The Geography of Private Sector Agricultural Innovation in the USA: Evidence from Patents

Matt Clancy

April 2023

**Abstract:** This paper develops a novel estimates of annual private sector agricultural R&D at the level of US states for the period 1976-2014. For each of five different agricultural subsectors, I allocate estimates of national private sector R&D across the 50 states by using the geographic distribution of inventors listed on contemporaneous US patents in the same agricultural subsector. These five subsectors comprise a large majority of total private sector agricultural R&D. I then use this new dataset to document three stylized facts about private sector agricultural R&D: it is highly correlated with the size of the state's agricultural economy (including over time, as well as in cross section), it is highly persistent, and is has become increasingly less concentrated over 1976-2014, though this last trend shows signs of reversing.

Affiliation: Coefficient Giving

Email: matt.clancy@coefficientgiving.org

American agriculture has experienced spectacular productivity growth over the last century, largely as a result of improved technologies and management practices that ultimately stem from research and development (R&D) (OECD 2016, Pardey and Alston 2021). R&D that results in productivity enhancing technologies in agriculture can have many origins, including explicitly agricultural R&D funded by the public sector, publicly funded science more generally, and non-agricultural R&D undertaken by the private sector (see Clancy et al. 2021). However, a very important locus is private sector agricultural R&D. Private sector agricultural R&D is where disparate ideas are integrated, finetuned, and transformed into viable products that can be sold in a market (see Graff, Rausser and Small 2003 for an application related to the GMO revolution). Accordingly, understanding both the determinants and the impacts of private sector agricultural R&D is important for understanding the broader system that generates agricultural productivity growth.

However, study of private sector agricultural R&D has long been stymied by data constraints. Whereas there exist long data series of public sector agricultural data at the level of individual states, such a dataset for private sector agriculture is absent. This absence becomes all the more pressing as an increasing share of agricultural R&D is conducted by the private sector. While the public sector has long been the dominant source of R&D related to the productivity of agriculture, the private sector's contribution is also non-trivial since at least the 1970s, and ultimately overtook the public sector as the primary source of agricultural R&D in the 2000s (Clancy, Fuglie and Heisey 2016).

The primary contribution of this chapter is to develop a new dataset estimating annual private sector agricultural R&D for individual US states over the period 1976-2014, and to document some basic stylized facts about the geography of private sector agricultural R&D. To construct this dataset, I marry two distinct datasets that have been independently developed. First, Fuglie et al. (2011) created long time series estimates for private sector agricultural R&D, but at the global and national level. However, importantly for our purposes, this dataset is broken down into major agricultural subsectors, for example, estimates of private sector R&D specific to the crop seed and biotech, crop protection chemicals, farm machinery, fertilizer, and animal health subsectors. This paper uses a US-specific update to Fuglie et al. (2011), provided by the Keith Fuglie, which brings data up to 2014.

The second dataset I use is the 1976-2016 dataset of agricultural patents, created in Clancy et al. (2021). The vast majority (98%) of agricultural patents belong to the private sector and so provide a parallel measure of private sector agricultural research effort. Importantly for this paper, these patents are also broken down into a set of subsectors with close analogues to the subsectors in Fuglie et al. (2011). Patents are also rich documents, and in this paper I exploit their data on the locations of inventors listed on patents.

The key idea of this paper is to assume that patents in Clancy et al. (2021) provide a good estimate of the *geographical distribution* of private sector agricultural R&D effort, while the national private R&D estimates of Fuglie et al. (2011) provide a good estimate of *total R&D spending* in each year. For each *subsector* of agricultural R&D, we distribute national R&D among the states using the contemporaneous geographic distribution of patents (as measured by the date a patent application is filed). We can then sum the R&D conducted by each subsector to estimate total private sector R&D conducted in each state, for each year.

Armed with this new dataset estimating private sector agricultural R&D across the fifty states over 1976-2014, this paper conducts an analysis to establish some initial stylized facts about the geography of agricultural R&D in the United States. I document three main facts.

First, in panel regressions, I show that private sector agricultural R&D is strongly correlated with other measures of a state's agricultural orientation. The most robust correlation is between the extent of agricultural R&D and the size of the agricultural economy, measured by the value of state agricultural production. I show this correlation persists, even when I include state-level fixed effects and national time-effects, so that within-state changes over time in the size of the agricultural economy are correlated with changes over time in the extent of agricultural R&D. In cross-section, private sector agricultural R&D is also strongly correlated with public sector agricultural R&D, but this correlation disappears when state fixed effects are included.

Second, I show that agricultural R&D is quite persistent. Between the 1980s and 2000s, the correlation of the log of average R&D across states is approximately 0.9. Knowing average agricultural R&D for a state in the 1980s is highly predictive of average R&D twenty years later.

Third, I show how the geographic concentration of private sector agricultural R&D has changed over 1976-2014. I show that, while private sector agricultural R&D is more concentrated than agricultural economic activity in general, this concentration has declined significantly from peaks in the early 1980s. This diffusion of R&D across the states has begun to reverse itself somewhat in the 2000s, but the extent of this reversal not universal and is less than for non-agricultural R&D. In general, agricultural R&D today is more diffuse than non-agricultural R&D (proxied by non-agricultural patents), though this has not always been the case.

The paper has three main sections. In section 1, I describe how the dataset is constructed. In section 2, I describe the panel regression results, which establish some broad correlates with state-level private sector agricultural R&D. Section 3 examines the evolution of private sector agricultural R&D across time, both in terms of its persistence and its geographic concentration. Finally, section 4 discusses the results and suggests some next directions for research.

## 1. Data Construction

The primary innovation of this paper is to use the locations of the inventors listed on subsector specific agricultural patents to apportion R&D across the US states. We begin by describing the construction of this patent dataset.

### 1.1. Identifying Agricultural Patents

A dataset of US patents for inventions with a primarily agricultural application was developed in Clancy et al. (2021). Full details of the construction of that dataset can be found in that paper. In brief, for five different agricultural subsectors, Clancy and coauthors use a combination of methods to identify relevant patents for agricultural technology. The primary methods relies on the US patent and trademark office's cooperative patent classification (CPC) system, which assigns patents to different technological categories based on the judgment of the patent examiner. Once suitable CPC codes can be identified, it is possible to identify all patents assigned these codes. In some cases, theses codes are not sufficiently detailed, in which case Clancy and coauthors supplement this CPC-based search system with manual inspection of patent documents. The five agricultural subsectors for which this approach is feasible are the biocides, fertilizer, machinery, plants, and agricultural research inputs subsectors.

To obtain data on animal health patents, Clancy and coauthors adopted a different approach based on Clancy and Sneeringer (2018). While the CPC system suffices to identify patents related to medical technology, it does not distinguish between medical technologies for human versus non-human animal application. Instead, to identify patents for veterinary medicine technologies, we rely on US Food and Drug Administration (FDA) archival data. Using archival records, Clancy and Sneeringer (2018) develop a list of all patents associated with approved veterinary medicine products.

It should be noted that the patents in the animal health subsector are subject to a selection effect that is not present in the other sectors. This is because animal health patents are only included if they are associated with veterinary drugs that eventually receive FDA approval. Drugs that are not approved may have associated patents, and we miss these. This selection effect may bias our results for this subsector in two ways. First, for most of our sample, we will miss any patents associated with veterinary products whose applications for FDA approval are ultimately unsuccessful. Accordingly, we observe only the subset of higher value patents associated with FDA-approved products. Second, towards the end of our sample, we face a truncation problem specific to the animal health sector. We do not observe patents associated with veterinary products that will ultimately be successful, but whose applications have not yet been approved. Since we date patents by their application date, it is likely that we increasingly understate the number of patents in animal health towards the end of our sample. Indeed, as seen in figure 1 below, the number of patents in animal health does exhibit an unusually steep decline towards the end of our sample period. However, as documented in Clancy and Sneeringer (2018), it is likely that this decline does not reflect only this double-truncation problem, since the actual number of approvals of FDA-approved drugs with novel ingredients also declined over this period, and these kinds of drugs are most likely to protected by patents.

Note this paper's results does not rely on each year containing an accurate patent count. However, our results do require the *share* of patents from each state to be representative of actual R&D conducted in the USA. We will return to this issue in a later section.

### 1.2. Mapping Patents to US States

Taking patents as signals of private sector R&D effort (Clancy et al. 2021 shows public sector patents accounted for only 2% of agricultural patents), we next need to locate this private R&D in geographical space. We used the addresses (city, state, and country) associated with the inventors listed on each patent to assign patents to states. This data is available from the patentsview website. Our focus is on research that occurs in the USA. For patents with some or all inventors residing abroad, we classify the patent as fractionally foreign, based on the share of its inventors that reside abroad (i.e., a patent with 1 out of 2 inventors residing abroad is listed as 0.5 foreign patents). By this measure, plants have the fewest foreign patents (14%) and biocides the most (49%). The rest of this paper focuses on the fraction of US patents associated with inventors who list US cities and states as their residence.

Where patents had more than one inventor belonging to different states, we fractionally allocated patents to different states. After linking patents to inventors and inventors to locations, we built a dataset of the number of patents applied for in each state, in each agricultural sector, in each year from 1976-2015. We use the year of a patent application, rather than its grant, to assign patents to different years, as this date more closely corresponds to the year R&D was conducted.

### 1.3. Aggregate Agricultural R&D Index

For each of 50 US state, this process yields a dataset consisting of 6 annual counts of (fractional) patents, running from 1976 to 2015. Each of these series corresponds to one of six different agricultural subsectors. My next step is to create a single index of aggregate private agricultural R&D for each state. Before describing my approach, I briefly discuss why a seemingly simpler alternative is undesirable: why not simply add up all the patents across subsectors to create annual state-level counts of agricultural patents?

This approach is inappropriate because as we will see, patents belonging to different subsectors and time periods are unlikely to represent commensurate levels of agricultural R&D. This is for a few reasons. First, subsectors may differ in their propensity to patent. In a 1994 survey of firms, Cohen, Nelson, and Walsh (2000) show there is wide variation across industries in their assessment of the value of patents in protecting inventions. Second, as noted above, our count of animal health patents is qualitatively different the other sectors, as it only identifies the subset of animal health patents associated with FDA approved veterinary products. Third, various legal and policy changes may have shifted the appeal of patents across time, such that a greater share of inventions may be patented in some periods than in others. Specifically, in the 1980s there was a large increase in the annual number of patents sought, and this may be the result of policy changes that made patenting more attractive, rather than because there was an increase in R&D effort (see Kortum and Lerner 1998 for discussion). Finally, towards the end of our dataset, we must increasingly contend with a well-known truncation issue in patent data. Patents are assigned to years based on their application date, but we only observe patents in our dataset if they have been granted at the time we download our data. Since it can take several years for patents to go from application to grant, in the last few years of our dataset, we see an increasingly small share of the patents that have been applied for and which will ultimately be granted. As noted above, this problem is likely to be more severe for the animal health subsector, where we face a double-truncation problem, since we additionally only observe patents if they are associated with FDA products approved by 2015.

Rather than add up patents of heterogeneous value to estimate total agricultural R&D, I rely on updated national estimates of US private sector agricultural R&D originally compiled as part of Fuglie et al. (2011) and apportion these headline numbers out among the states based on the share of patents from each state. These updated estimates are helpfully broken down into agricultural subsectors with a close correspondence to our agricultural patent datasets, covering the same period as our patents up to 2014. These estimates are at the national level, but I assume the geographic distribution of patents is reflective of the contemporaneous geographic distribution of R&D funds. So long as there are no systematic differences between where R&D is conducted and which R&D results in a patent, this will provide an unbiased estimate of where private sector R&D occurs.

Table 1 displays the correspondence used in this paper between the subsectors in Fuglie et al. (2011) R&D estimates and the subsector for which we have state-level patent counts. We also display the share of all private agricultural input R&D accounted for by each subsector's private R&D spending.

**Table 1. R&D – Patent Correspondences**

| Private Sector R&D from Fuglie et al. (2011) | Patent Series | Share of Total Private Ag Inputs R&D |
|---|---:|---:|
| Crop protection chemicals | Biocides | 32% |
| Crop seed & Biotech | Plants + Research | 27% |
| Farm Machinery | Machinery | 24% |
| Animal Health | Animal Health | 17% |
| Animal Genetics | - | 5% |
| Animal Nutrition | - | 3% |
| Fertilizers | Fertilizer | 1% |

We make three observations about Table 1. First, note that we combine patents from the plant and agricultural research inputs subsectors, and use this combined series to apportion R&D spending related to crop seed and biotechnology. This is because crop seed and biotechnology R&D funding is not easy to conceptually break down into R&D related to plants and R&D related to breeding and genetic modification technologies. Second, note our approach will not work for the animal genetics and animal nutrition subsectors, as these subsectors either do not rely on patents for protection, or we do not have a reliable way of identifying their patents. However, as is clear from the third column of Table 1, these subsectors account for just a small share of overall private sector agricultural R&D. Finally, attentive readers may also notice that the numbers in column 3 add up to 109%. This is because the animal health R&D figures we use include R&D on veterinary products for agricultural and non-agricultural animals, which matches our patent data.

### 1.4. Assessing the Match Between Series

Before proceeding, figure 1 illustrates the correspondence, over time, between our patent data, when aggregated up to the national level, and these national R&D estimates. In each of the five figures, the dashed line corresponds to the 5-year moving average of Fuglie et al. (2011)'s private agricultural R&D series. The solid line corresponds to 5-year moving average of US patents by subsector.

**Figure 1. Real Private Sector R&D vs National Patent Counts by Sub-Sector, 5-year moving average**

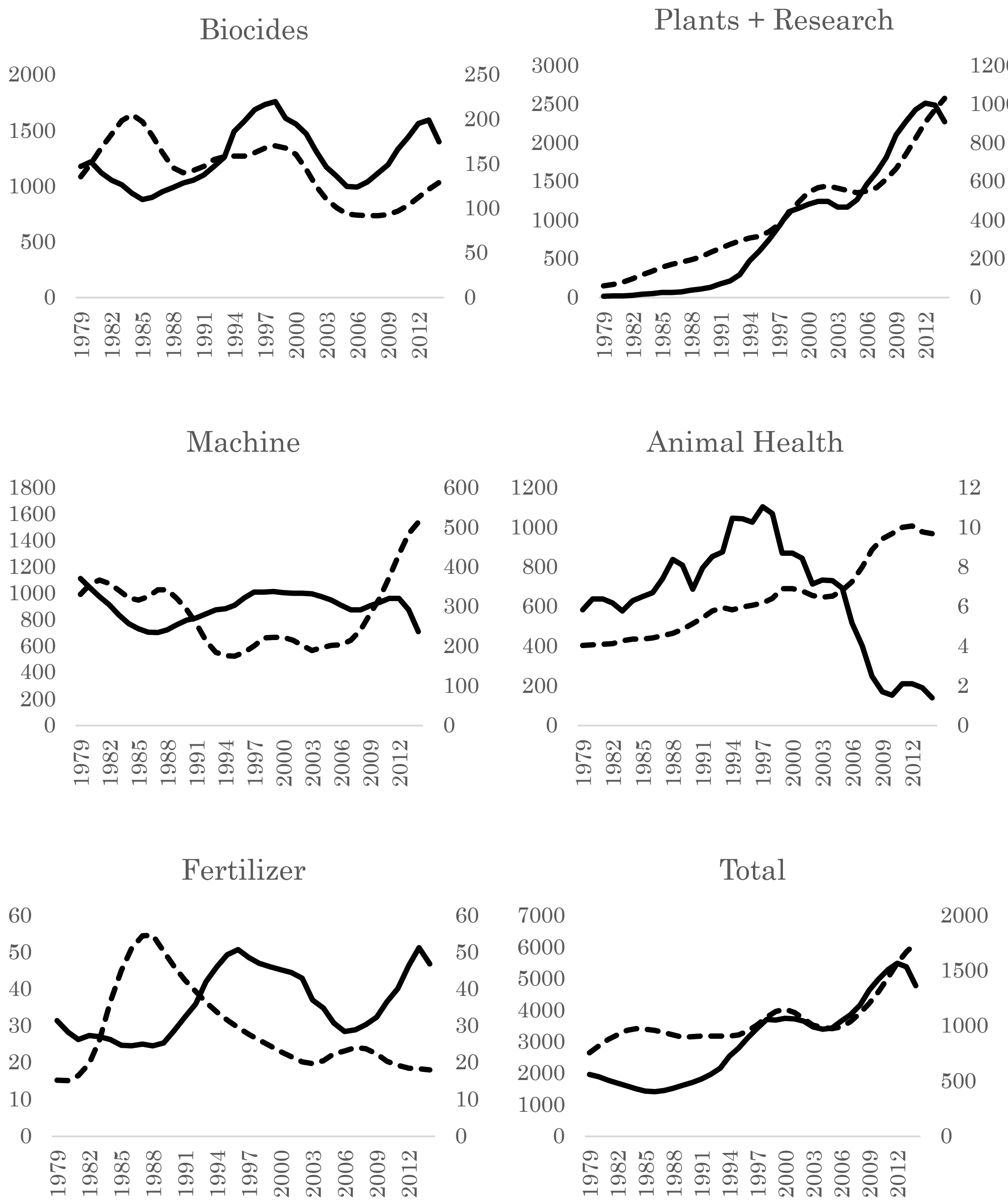

Notes: Solid Line corresponds to 5-year moving average of US patent counts (right axis). Dashed line corresponds to 5-year moving average of private sector agricultural R&D in 2013 million dollars (left axis). Total is a count of all agricultural patents and all private agricultural R&D.

A few observations are in order.

First, while there is some general correspondence for some of these subsectors, overall the correspondence is quite poor. The fit is best for plants, which account for 27% of R&D spending on ag inputs. Second, note that each patent series dips at the end of the sample – this is the afore-mentioned patent truncation issue. This dip is more severe for animal health which suffers from a potential double truncation issue, though as discussed earlier, this dip is likely not entirely illusory.

Third, the plants and research inputs subsectors exhibit a sharp increase from zero (or close to zero) in the 1980s. This is due to legal changes in the patentability of biological innovation in the wake of the 1980 Diamond v. Chakraborty Supreme Court case (Clancy and Moschini 2017). Prior to 1980, biological innovations such as new plant varieties were not patentable subject matter. It is important to note that R&D related to biological innovation that occurs prior to 1980 may be understated in the patent record, since we count relevant research patents, but not patents for plants.

Finally, note that subsectors vary significantly in the size of their patent and R&D portfolios. Figure 2 plots the number of patents per year by sector on the horizontal axis and the number of R&D dollars, in 2013 millions along the vertical axis. Given the large differences in the size of sectors, figure 2 is plotted in a log scale. We also exclude data post-dating 2010, to minimize the truncation problem. Figure 2 makes clear that, even if year-to-year patenting is only weakly correlated with year-to-year R&D spending, across sectors the correlation is much stronger.

**Figure 2. Real Private Sector R&D vs National Patent Counts by Sub-Sector, 1976-2010**

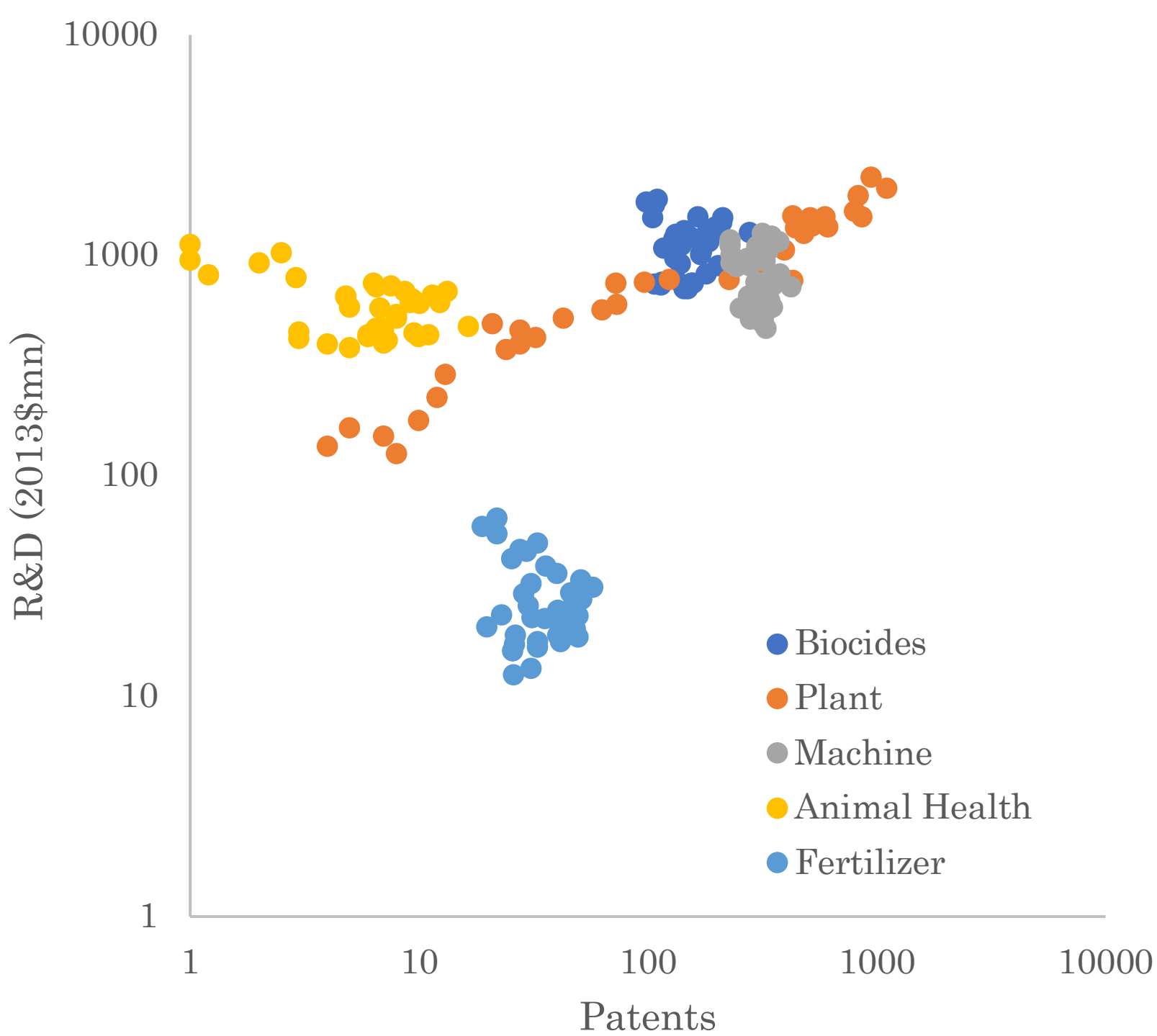

The correlation coefficient between annual R&D spending and patents, across all sectors is 0.63. If we exclude the animal health subsector, for which we observe only a subset of patents, the correlation rises to 0.67. And in log terms, as in figure 2 (but again excluding animal health), the correlation rises to 0.73.

These two figures suggest that patents are reasonably correlated with R&D in cross-section, but only very weakly correlated within any sector over time. My approach to creating an aggregate R&D index is to use the private R&D estimates of Fuglie et al. (2011), but to allocate this spending across states based on the contemporaneous share of patents belonging to each state. Formally, define our private R&D estimate $X_{st}$ for state $s$ in year $t$ as:

$$X_{st} = \sum_i \frac{p_{ist}}{\sum_s p_{ist}} Y_{it}$$

Where $p_{ist}$ is the count of patents belonging to subsector $i$ in year $t$ in state $s$ and $Y_{it}$ is the total private R&D spending in subsector $i$ in year $t$.

Thus, our results do not depend on the match between the absolute count of patents year-to-year and R&D dollars in that year, but instead on the assumption that the geographic distribution of patents in any given year is a good proxy for the geographic distribution of R&D conducted in the same year. Nonetheless, figure 1 suggests we should be very cautious about interpreting year-to-year changes in this index. Accordingly, this paper emphasizes broad patterns and averages.

### 1.5. Data Description

We now have 7 data series for each state: one counting annual patents for each subsector, and our aggregate R&D index, which estimates total private sector agricultural R&D performed in the state in that year (measured in millions of 2013 dollars). This results in 40 years of data, for 50 different states, times 7 series. Table 2 presents some summary statistics for this data series.

**Table 2. Summary Statistics**

| | Animal Health | Biocide | Fertilizer | Machine | Plant | Research | Aggregate |
|---|---|---|---|---|---|---|---|
| Mean | 0.23 | 5.65 | 1.32 | 10.8 | 7.54 | 5.86 | 136 |
| Median | 0 | 1 | 0 | 2.33 | 0 | 0.33 | 28.2 |
| Max | 16.5 | 278 | 67.2 | 424 | 913 | 381 | 6352 |
| St. Dev. | 1.05 | 21.7 | 5.15 | 40.8 | 46.9 | 28.3 | 517 |

Note that in every subsector, private sector R&D is quite skewed, with the mean well above the mean in every case, and the standard deviation significantly above the mean.

## 2. Correlates of Private Agricultural R&D

Our first application of this data is to establish some broad cross-sectional correlates of private agricultural R&D at the state level. The following analysis is not intended to pin down the causal determinants of private sector R&D, but merely to establish some statistical associations.

We investigate two main correlates: the size of the state agricultural economy, and the level of state *public* sector agricultural R&D. For state-level data on the size of the state agricultural economy, we use the real value of agricultural sector production time series from the USDA/ERS Farm Income and Wealth statistics.[1] For state level real public R&D spending, we use data from Jin and Huffman (2016) which ends in 2009. Jin and Huffman is missing data Alaska and Hawaii, which reduces our sample from 50 to 48 states.

We estimate variations of the following simple OLS regression:

[1] https://data.ers.usda.gov/reports.aspx?ID=17830

$$\log X_{st} = \beta_1 \log(Value\ Ag\ Production)_{st} + \beta_2 \log(Public\ R\&D)_{st} + \alpha_t + \gamma_s + e_{st}$$

Where $X_{st}$ is our state-level estimate of private sector R&D. The index *s* refers to states, while *t* initially refers to a *decade*: 1980-1989, 1990-1999, and 2000-2009. Each variable takes on the average value over this decade. We initially collapse our observations from annual to decadal for two reasons. First, as noted above, we have significant year-to-year uncertainty about the exact allocation of private R&D across states. Second, in 6.1% of our year-state observations, our private index of R&D is equal to zero which complicates estimation in logs. When we collapse down to averages over a decade, we no longer have any dependent variables equal to zero. We will also estimate this equation with annual data, as discussed shortly.

The results of this decadal regression are given in table 3. Columns 1-3 exclude state-specific fixed effects, and columns 4-6 add them. Our main conclusion from this exercise is that states with a greater agricultural orientation have more private sector agricultural R&D. Specifically, in column (1), when we include only the value of agricultural production, across states we find a 10% increase in value of agricultural production is associated with a 7.5% increase in private agricultural R&D. In column (2), when we switch this for the value of public sector agricultural R&D, we find an even stronger relationship: private agricultural R&D increases one for one with public R&D, when looking across states in a given decade. Note that we always include time-fixed effects, to control for broad changes in agricultural R&D. As indicated by the dashed line in “Figure 1: total”, total agricultural R&D is significantly higher over 2000-2009 than in other decades.

**Table 3. Correlates of Private Agricultural R&D by Decade**

| | (1) | (2) | (3) | (4) | (5) | (6) |
|---|---|---|---|---|---|---|
| | | | | | | |
| Value of Ag Production | 0.751*** | | 0.061 | 0.637 | | 0.614 |
| | (0.133) | | (0.272) | (0.781) | | (0.748) |
| Public R&D | | 1.000*** | 0.931** | | 0.121 | 0.073 |
| | | (0.151) | (0.307) | | (0.276) | (0.251) |
| Fixed Effects | Time | Time | Time | Individual + Time | Individual + Time | Individual + Time |
| Observations | 144 | 144 | 144 | 144 | 144 | 144 |
| R-squared | 0.390 | 0.456 | 0.457 | 0.011 | 0.002 | 0.012 |

Notes: White standard errors, clustered by state, are given in parentheses. Statistical significance levels are indicated by asterisks, where *** = 0.1%, ** = 1%, * = 5%

The value of agricultural production and public agricultural R&D are themselves highly correlated, with a 0.91 pearson correlation coefficient in logs. Unsurprisingly, when we include both of them in the regression model of column 3, only one is significant. Similarly, when we add state-specific field effects, as in columns 4-6, nothing is significant anymore. This suggests the results of columns 1-3 are picking up each states' broad long-run orientation towards agriculture. Decade-to-decade fluctuations in public R&D or the size of the local agricultural economy have no statistically significant correlation with private agricultural R&D, once each states' average tendency to invest in private sector agricultural R&D is taken into account.

In tables 4 and 5, we re-run these models with annual, rather than decadal data. Because, 6.1% of our observations have $X_{st} = 0$, we use two strategies to estimate our log specification. In table 4, we drop all observations with $X_{st} = 0$, while in table 5 we retain these observation but change the dependent variable to $\log(1 + X_{st})$. In both cases we obtain very similar results to table 4 for columns 1-3.

However, in columns 4-6, with more observations we are now able to detect a statistically significant relationship between the size of the agricultural economy and contemporaneous private R&D expenditures, even when we add in state-specific fixed effects. These effects are comparable in magnitude to the magnitudes detected in columns 1, when state-fixed effects are not added. We continue to fail to reject the null hypothesis that there is no correlation between contemporaneous private and public R&D, once state-level fixed effects are included. We discuss our interpretation of these results in the conclusion.

**Table 4. Correlates of Private Agricultural R&D by Year, dropping observations with $X_{st} = 0$**

| | (1) | (2) | (3) | (4) | (5) | (6) |
|---|---|---|---|---|---|---|
| | | | | | | |
| Value of Ag Production | 0.717*** | | 0.206 | 0.697* | | 0.680* |
| | (0.130) | | (0.249) | (0.781) | | (0.301) |
| Public R&D | | 0.925*** | 0.701** | | 0.198 | 0.176 |
| | | (0.144) | (0.271) | | (0.158) | (0.156) |
| Fixed Effects | Time | Time | Time | Individual + Time | Individual + Time | Individual + Time |
| Observations | 1487 | 1487 | 1487 | 1487 | 1487 | 1487 |
| R-squared | 0.301 | 0.336 | 0.341 | 0.01 | 0.003 | 0.012 |

Notes: White standard errors, clustered by state, are given in parentheses. Statistical significance levels are indicated by asterisks, where *** = 0.1%, ** = 1%, * = 5%

**Table 5. Correlates of Private Agricultural R&D by Year, dependent variable is log(1 + $X_{st}$)**

| | (1) | (2) | (3) | (4) | (5) | (6) |
|---|---|---|---|---|---|---|
| | | | | | | |
| Value of Ag Production | 0.789*** | | 0.218 | 0.668* | | 0.668* |
| | (0.119) | | (0.238) | (0. 274) | | (0.0.275) |
| Public R&D | | 1.002*** | 0.781** | | 0.020 | -0.002 |
| | | (0.133) | (0.270) | | (0.276) | (0.146) |
| Fixed Effects | Time | Time | Time | Individual + Time | Individual + Time | Individual + Time |
| Observations | 1584 | 1584 | 1584 | 1584 | 1584 | 1584 |
| R-squared | 0.372 | 0.412 | 0.417 | 0.011 | 0.000 | 0.011 |

Notes: White standard errors, clustered by state, are given in parentheses. Statistical significance levels are indicated by asterisks, where *** = 0.1%, ** = 1%, * = 5%

3. Changing Geography of Agricultural Innovation

Our second application leverages the time dimension of our data to look for broad changes in the geography of private agricultural R&D. We begin by assessing how persistent is private agricultural R&D by comparing our estimates of private R&D over the period 1980-1989 to the period 2000-2009. Figure 1 plots the log of average private R&D by state over 1980-1989 on the horizontal axis and log of average private R&D by state twenty years later on the vertical axis. The two are strongly correlated, with a correlation coefficient of 0.87 in logs.

**Figure 3. Log Average Agricultural R&D Spending, 1980s vs. 2000s.**

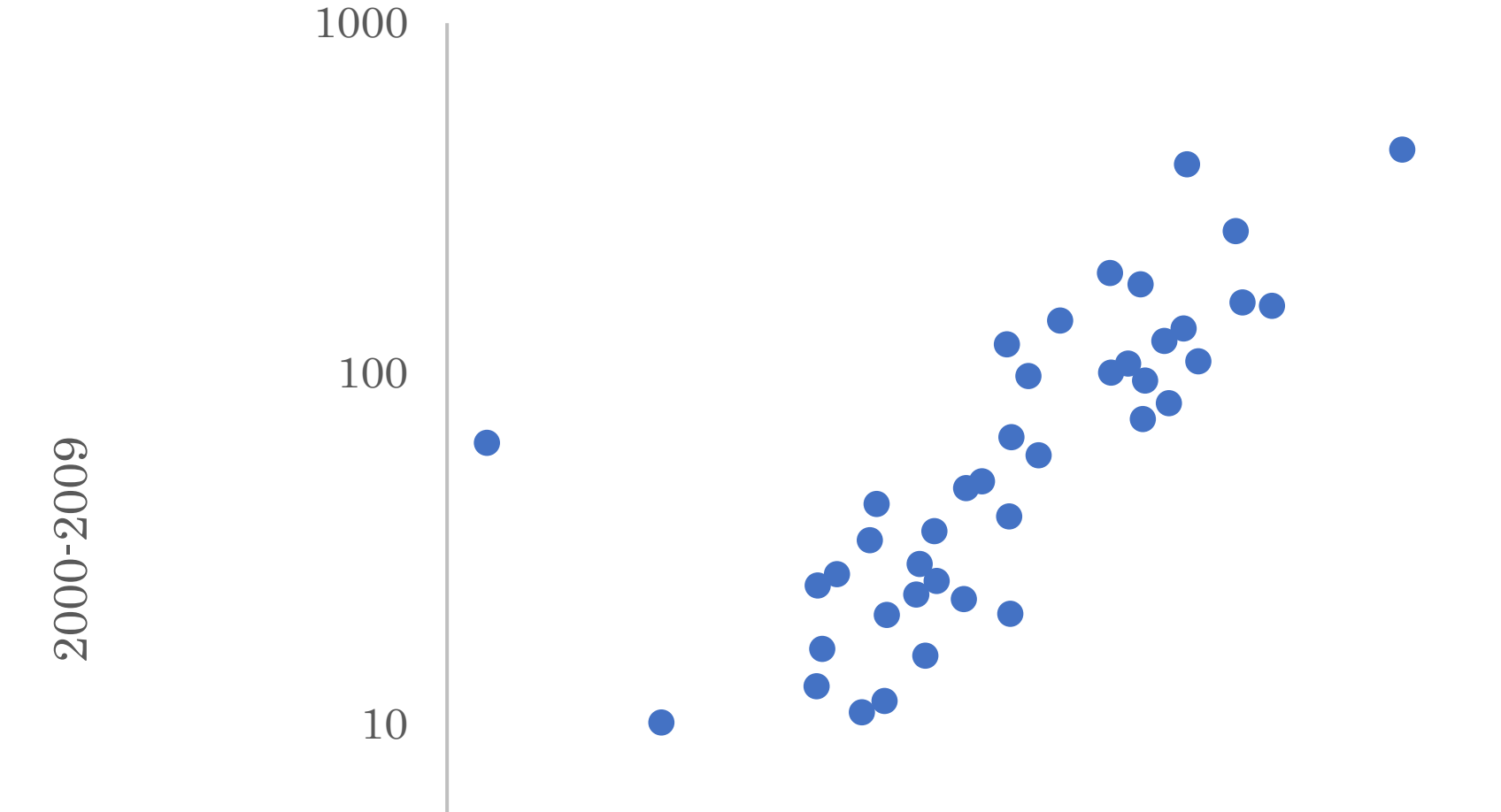

The largest outlier is Maine. We estimate Maine's private R&D averaged slightly over $1mn per year during 1980-1989, and slightly under $64mn per year over 2000-2009. But this result is quite likely to be overstated. It is primarily driven by 7 patents awarded to Maine-based inventors in animal health in 2003, 2004, and 2009. Patenting in animal health began to drop significantly in the 2000s, and these 7 patents accounted for 40% of animal health patents during these three years. At the same time as aggregate patenting in this sector began to fall, R&D began to rise significantly, so that Maine is allocated 40% of a large animal health R&D spend in these three years. This includes 40% of animal health's near peak R&D spending in 2009, since Maine has 1 of the 2.5 US animal health patent applications in that year. Excluding those three years, Maine's average R&D index over 2000-2009 would be $11.2mn. Notably, Maine would still remain an outlier, though no longer as extreme a one. Excluding Maine from our data, the correlation across time period is 0.93 in logs.

Another way to measure the geographic evolution of agricultural R&D is to compute the Herfindahl concentration index of private R&D in agriculture, as in figure 4. The Herfindahl concentration index is typically defined as the sum of the squared shares of some quantity, and so a natural implementation in this case would be to compute the annual share of patents belonging to each state, square them, and sum. However, given the imprecision in our year-to-year estimates of state-level R&D, in figure 4, we use data from the preceding five years, instead of only the current year. The Herfindahl concentration index from figure 4 is defined here as:

$$H_t = \frac{\sum_s \left(\sum_{t-4}^{t} X_{st}\right)^2}{\left(\sum_s \sum_{t-4}^{t} X_{st}\right)^2}$$

In other words, this index is the square of each state's R&D (or count of sector-specific patents) from the preceding *five* years, divided by the square of the total R&D (or sector-specific patents) over the same time period. A higher number indicates R&D is more concentrated in a smaller number of states, up to a max of 1, which would indicate all agricultural R&D is performed in one state, and a minimum of 0.02, which would indicate agricultural R&D is evenly divided among 50 states.

**Figure 4. Herfindahl Concentration Index of lagging 5 years, Ag Patenting and Private R&D Index (solid black), Non-Ag Patents (solid gray), Value of Ag Production (dotted black)**

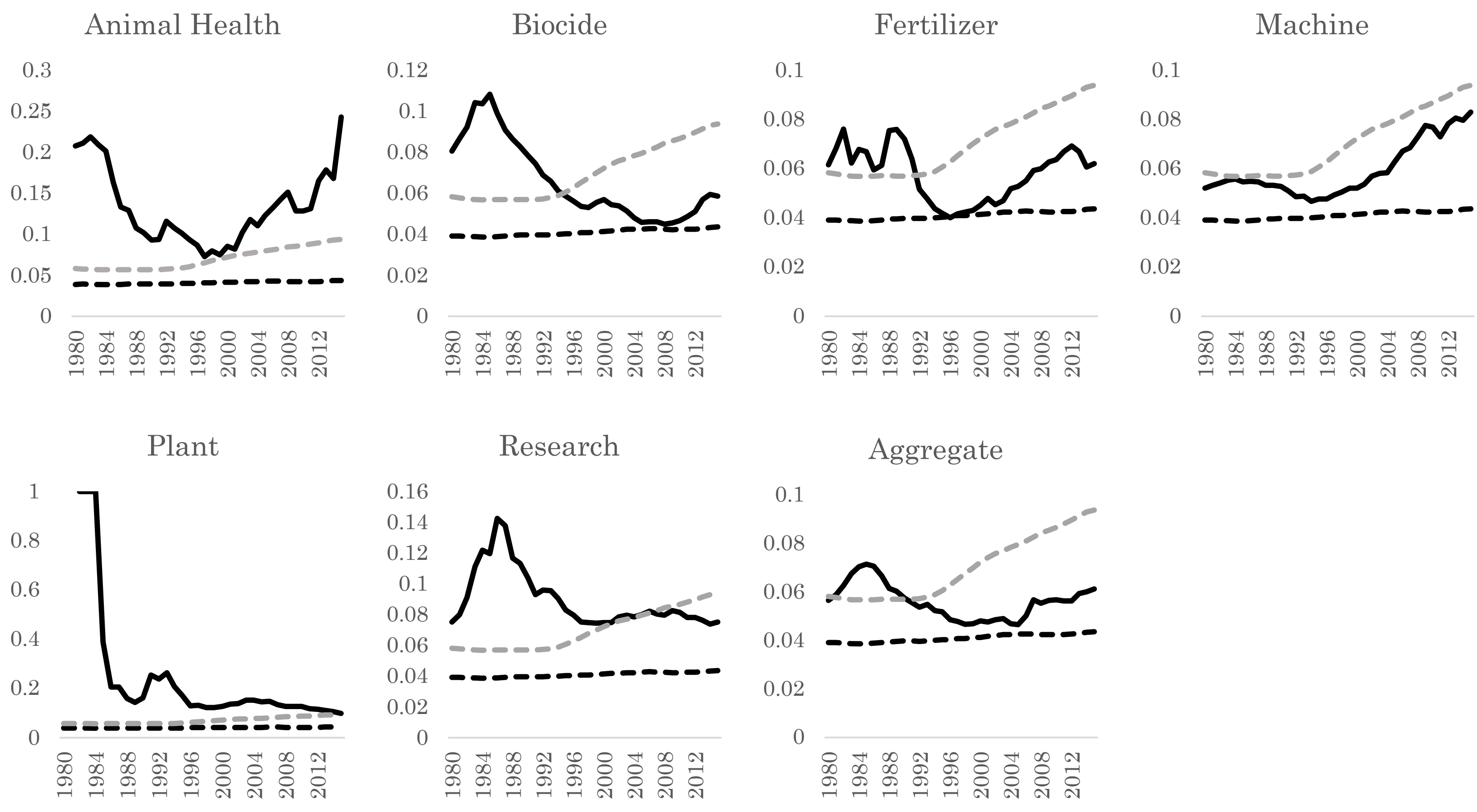

As a comparison, figure 4 also includes two additional concentration indices, each of which is also computed based on the preceding five-years of data. The black dashed line is the concentration index for the value of agricultural production in the states. The gray dashed line is the concentration index for all *non-agricultural* patents in the United States. It is also possible to compute the concentration of public sector R&D through 2009, but this line very closely follows the trend for the value of agricultural production.

We can make a few observations. First, the solid black line lies above the dashed black one at nearly all points, indicating in aggregate and at the level of each subsector, private agricultural R&D tends to be more concentrated than agricultural production as a whole. Second, while the concentration of agricultural R&D barely budges, the geographic concentration of R&D exhibits more.

Our concentration measure of private agricultural R&D indicates that agricultural R&D was initially more concentrated than non-agricultural R&D (as measured by patents), but while non-agricultural R&D was broadly stable through 1990, the concentration of agricultural R&D began declining in the 1980s and fell below non-agricultural levels in the 1990s. Since the 1990s, non-agricultural patenting has become increasingly concentrated in a smaller number of states (a finding also found at the county-level in Andrews and Whalley 2021), and agricultural R&D began to exhibit the same trend though at a more muted level and with a considerable lag.

This rebound in concentration should be interpreted with some caution though. The animal health sector, for example, accounts for 16% of our aggregate index on average, and the number of patents declined sharply over this period (see "Figure 1: Animal Health"). This will mechanically generate an increase in concentration, since it can be shown the minimum value a concentration index can attain is the reciprocal of the number of patents we observe. Notably, the animal health subsector does exhibit the sharpest increase in concentration. That said, the rebound in concentration is probably real, though possibly overstated given uncertainty about our animal health data. The machine and fertilizer subsectors exhibit sustained increases in concentration over this period, and collectively account for a greater share of the index than animal health. Neither do they suffer from a precipitous decline in the number of patents we observe over this period.

Finally, we note that the rapid decline in concentration observed in the plant subsector is also a mechanical artifact of the increase from very small number of patents in that subsector prior to 1990. The research inputs subsector has some similar issues, though not as extreme as in the plants subsector. Note, however, that in our aggregate index these sectors are given relatively lower weight in these years because private R&D spending in the sector was also relatively low at this time (see "Figure 1: Plants + Research").

## 4. Conclusion and Discussion

This work is a first step in understanding the dynamics of the geography of private sector agricultural R&D in the United States. It constructs a novel database of state-level private agricultural R&D by merging agricultural subsector specific datasets on private R&D and patents and using the distribution of inventors on patents to infer the distribution of private R&D funding.

Taken together, this paper suggests a private agricultural R&D system whose geography changes slowly. The correlation across states between R&D in the 1980s and the 2000s is approximately 0.9; states that performed a lot of agricultural R&D in the 1980s still perform a lot in the 2000s. This initial and slowly changing distribution of private agricultural R&D appears to be closely connected to the agricultural orientation of each state. In any given decade, for example, states with a larger agricultural economy, or whose governments spend more on agricultural R&D, also have correspondingly larger agricultural R&D. Reflecting the relative inflexibility of agricultural R&D, once we account for each state's time invariant characteristics by including fixed effects, decadal changes in public sector R&D or the size of the local agricultural economy have no impact on fluctuations in agricultural R&D, over a thirty-year time horizon.

But slow change is not the same as no change. This paper documents changes in two dimensions. First, when we use annual rather than decadal data, we find there continues to be relationship between the contemporaneous size of a state's agricultural economy and private sector funding, even when we account for state fixed effects. In other words, states whose agricultural economy grows at a slower rate than the average also experience slower growth in their private sector agricultural R&D. Just so for states whose economy grows more quickly than the average. In contrast, we never are able to detect a statistically significant relationship between public and private agricultural R&D, once we include state-level fixed effects - which presumably capture time invariant orientation towards agriculture, which is likely correlated with public sector agricultural R&D.

While this paper lacks a causal identification strategy, the robust positive correlation between the size of the market and the extent of private R&D is consistent with a variety of work that has found causal linkages between the size of the market and R&D. This relationship could have multiple roots. It may be that R&D is more attractive when the market is larger (Acemoglu and Linn 2004, Duboi et al. 2015). Alternatively, given liquidity constraints, it may be that a larger market results in more funding for R&D, for example, via increased profits for firms (Krieger, Li, and Papanikolaou 2021).

Second, as measured by the Herfindahl concentration index, we observe a tendency for agricultural R&D to spread out over a greater number of states over the forty years for which we have data. This tendency is somewhat reversed in the 2000s, following broader trends in R&D outside of agriculture, but in aggregate the reversal is much smaller than the trends observed in the non-agricultural economy.

This paper is a first effort and it is my hope that this dataset opens up new opportunities for further work, both in understanding the determinants and the impact of private sector agriculture. For example, a long-running question in agricultural economics has been the extent to which public sector R&D is a complement or substitute for private sector R&D. This dataset could provide new ways of testing for such a relationship, by creating scope for identifying more localized effects. Turning to the impacts of agricultural R&D, there is a large literature (see Fuglie et al. 2016 for some discussion) that links state-level public sector R&D to state-level productivity and this dataset could enable a similar strategy for the private sector.